\documentclass[aps,pra,reprint]{revtex4-1}
\usepackage{amsmath}
\usepackage{amsfonts}
\usepackage{amssymb}
\usepackage{mathtools}
\usepackage{graphicx}
\usepackage[caption=false,position=b]{subfig}
\usepackage[colorlinks]{hyperref}
\usepackage[capitalize]{cleveref}

\usepackage{yypreamble}

\usepackage{color}
\usepackage[normalem]{ulem}

\newcommand{\C}{\hat{\mathcal C}}
\newcommand{\hS}{\hat{\mathcal S}}
\newcommand{\vm}{\vec{m}}
\newcommand{\vn}{\vec{n}}

\newcommand{\LL}{N}
\newcommand{\Ns}{\mathcal{N}}
\newcommand{\Ms}{\mathcal{M}}
\newcommand{\pardash}[1]{\section{#1}}

\begin{document}

\title{Reservoir engineering of bosonic lattices using chiral symmetry and localized dissipation}
\author{Yariv Yanay}
\affiliation{Laboratory for Physical Sciences, 8050 Greenmead Dr., College Park, MD 20740}
\author{Aashish A. Clerk}
\affiliation{Institute for Molecular Engineering, University of Chicago, 5640 S. Ellis Ave., Chicago, IL 60637}

\date{\today}

\begin{abstract}
We show how a generalized kind of chiral symmetry can be used to construct highly-efficient reservoir engineering protocols for bosonic lattices.  These protocols exploit only a single squeezed reservoir coupled to a single lattice site; this is enough to stabilize the entire system in a pure, entangled steady state.  Our approach is applicable to lattices in any dimension, and does not rely on translational invariance.  We show how the relevant symmetry operation directly determines the real space correlation structure in the steady state, and give several examples that are within reach in several one and two dimensional quantum photonic platforms.  
\end{abstract}

\maketitle

Pure quantum states with non-classical properties such as entanglement or squeezing play an important role in quantum computing and communication, and robust methods for their preparation are an important resource. One powerful general approach is reservoir engineering \cite{Poyatos1996,Plenio2002}, where carefully tailored dissipation is used to prepare and stabilize non-trivial quantum states.  Reservoir engineering of a few degrees of freedom is by now a well-established technique, and has been implemented experimentally in a range of systems spanning atomic physics, quantum optics, superconducting circuits and optomechanics  (see e.g.
Refs.~\onlinecite{Krauter2011,Murch2012,Wineland2013,Shankar2013,Leghtas2015,Wollman2015}).

Dissipative state stabilization methods can also be formulated for lattice systems having many degrees of freedom, potentially allowing the preparation of correlated and even topological states \cite{Diehl2008,Verstraete2008,Kraus2008,Cho2011,Koga2012,Ikeda2013,Quijandria2013,Ticozzi2013}.  Such proposals are typically resource intensive:  they usually require independent engineered reservoirs at every site or highly non-local dissipators that are difficult to construct. 
Given this, attention has recently focused on methods employing just a single, localized engineered reservoir to stabilize pure correlated states of a lattice. 
Previous studies have focused on one-dimensional (1D) systems, and considered specific examples \cite{Zippilli2015}, as well as general parameterizations of achievable steady states \cite{Ma2016,Ma2016a, Ma2017}.  
Despite this impressive work, simple physical principles determining when such a local reservoir engineering approach is possible are lacking, as is treatment of higher-dimensional systems.

In this Letter, we address this problem.  We demonstrate how symmetry can be a powerful tool for engineering a wide range of systems where a single, locally coupled reservoir is able to stabilize a non-trivial pure quantum state of a lattice.  We focus on a lattice of bosonic sites coupled to a squeezed reservoir at just a single site.  We show that if the lattice possesses a generalized chiral symmetry and no dark modes, the local reservoir relaxes the system into a pure steady state with a non-zero density, and correlation and entanglement properties directly related to the nature of the symmetry, and not dependent on further details of the system's eigenmodes.  

The understanding of the steady state in terms of the symmetry of the non-dissipative lattice Hamiltonian paves the way for custom design of lattice systems to be used in the preparation of a variety of many-mode steady states, as we demonstrate in several examples. 
Such states are particularly useful in continuous-variable and one-way quantum computation \cite{Weedbrook2012}.
  The results we present are directly applicable to a number of experimental platforms. In particular, experiments in superconducting circuits have recently demonstrated all the required elements for our approach, including the construction of non-trivial lattice structures  \cite{Underwood2012,Fitzpatrick2017,Owens2017} and the ability to couple strongly to a squeezed vacuum reservoir \cite{Flurin2012,Murch2013,Toyli2016,Clark2016}.

%%%%%%%%%%%%%%%%%%%%%%%%%%%%%%%%%%%%%%%%
%%%%%%%%%%%%%%%%%%%%%%%%%%%%%%%%%%%%%%%%
\begin{figure}[tbp] %  figure placement: here, top, bottom, or page
   \centering
   \includegraphics[width=0.75\columnwidth]{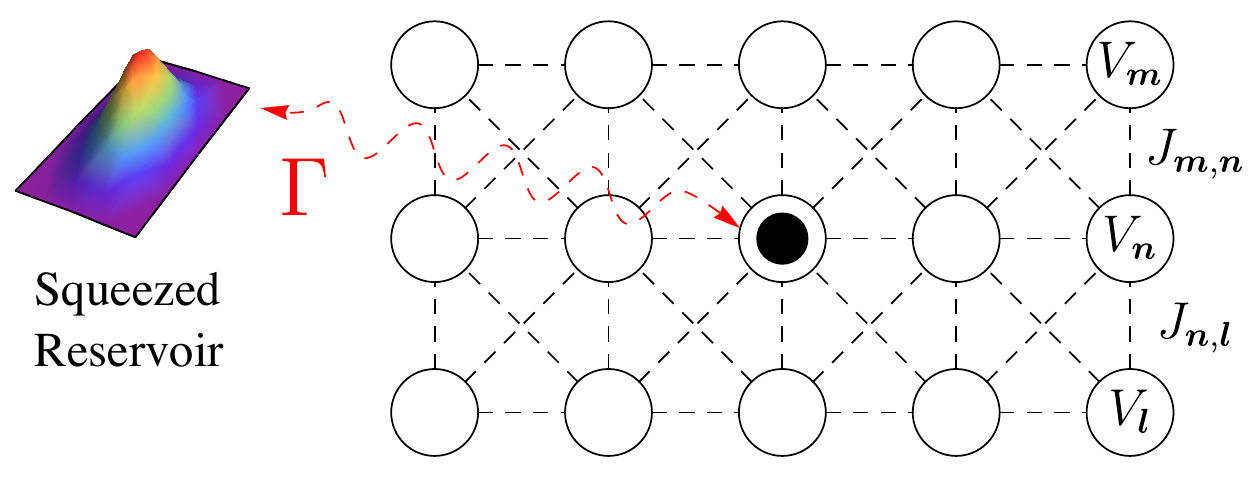} 
   \caption{Schematic of the system:  bosons hop between sites $\vn$ of an arbitrary lattice, as described by a Hamiltonian with hopping matrix elements $J_{\vm,\vn}$ and on-site potentials $V_{\vn}$.  One site of the lattice (denoted $\vn_0$, marked in black) is coupled to a squeezed vacuum reservoir (giving rise to a local damping rate $\Gamma$).
    }
   \label{fig:diagram}
\end{figure}
%%%%%%%%%%%%%%%%%%%%%%%%%%%%%%%%%%%%%%%%
%%%%%%%%%%%%%%%%%%%%%%%%%%%%%%%%%%%%%%%%

%%%%%%%%%%%%%%%%%%%%%%%%%%%%%%%%%%%%%%%%%%%%%%%%%%%%%%%%%%%%%%%%%%%
\pardash{Model}We start by considering bosons hopping on an arbitrary $d$-dimensional lattice of sites (see \cref{fig:diagram}), 
as described by a generic tight-binding Hamiltonian:
\begin{equation}\begin{split}
\hat \H & =  \sum_{\mathclap{\vm, \vn}}H_{\vm,\vn}\hat a_{\vm}\dg\hat a_{\vn} \\ & =  \sum_{\vn}V_{\vn}\hat a_{\vn}\dg\hat a_{\vn} + \sum_{\mathclap{\vm\ne \vn}}J_{\vm,\vn}\hat a_{\vm}\dg\hat a_{\vn}
\label{eq:HS}
\end{split}\end{equation}
Here $\hat a_{\vn}$ ($\hat a_{\vn}\dg$) is the annihilation (creation) operator for a boson on site $\vn$,  $J_{\vm,\vn} = J_{\vn,\vm}^{*}$ is the hopping strength between sites $\vm,\vn$, and $V_{\vn}$ is the potential on site $\vn$. Summations are over all $\LL$ sites in the lattice.  Note that we do not assume translational invariance.  

We take a single ``drain'' site, $\vn_{0}$, to be linearly coupled to a squeezed zero-temperature Markovian reservoir.  In a photonic realization of our system, where each site is a cavity, this simply corresponds to driving site $\vn_{0}$ with broadband squeezed vacuum noise.  
We use the standard input-output treatment of the resulting dissipation \cite{Gardiner2004}, yielding the Heisenberg-Langevin equations of motion
\begin{equation}
	\dot{\hat a}_{\vn} = -i\br{\hat a_{\vn},\hat \H} - \gd_{\vn,\vn_{0}}\mat{\half[\Gamma]\hat a_{\vn_{0}} - \sqrt{\Gamma}\hat \zeta}.
\label{eq:evolreal}
\end{equation}
The rate $\Gamma$ parameterizes the strength of the coupling to the reservoir and the operator $\hat \zeta$ describes the squeezed vacuum fluctuations associated with it.  This is operator-valued Gaussian white noise, with correlators
\begin{equation}\begin{gathered}
\avg{\hat \zeta\dg\p{t}\hat \zeta\p{t\pr}} = \gd\p{t-t\pr} \Ns, \quad \avg{\hat \zeta\p{t}\hat \zeta\p{t\pr}} = \gd\p{t-t\pr} \Ms,
\\ \Ns = \sinh^{2}r, \qquad \Ms = e^{i\phi}\cosh r \sinh r,
\end{gathered}\end{equation}
where $r$ ($\phi$) is the squeezing parameter (angle).

Diagonalizing our tight-binding Hamiltonian yields
\begin{equation}
\hat \H = \sum_i \gve_{i}\hat b_{i}\dg\hat b_{i}, \qquad \hat b_{i}\dg = \sum_{\vn}\psi_{i}\br{\vn}\hat a_{\vn}\dg,
\end{equation}
where $\hat{b}_{i}$ annihilates an energy eigenmode with mode energy $\gve_i$ and wavefunction $\psi_{i}\br{\vn}$.
Without loss of generality, we label the modes so that $\gve_{i+1}\ge \gve_{i}$. Including the coupling to the reservoir, the equations of motion in the energy eigenmode basis take the form
\begin{equation}
	\dot{\hat b}_{i} = -i \sum_{j} A_{i,j}\hat b_{j} + e^{-i\varphi_{i}}\sqrt{\bar \Gamma_{i}}\hat\zeta, 
\label{eq:evolbe}
 \end{equation}
where $\varphi_i = \arg \left( \psi_i\br{\vn_{0}} \right)$, $\bar \Gamma_{i} = \abs{\psi_{i}\br{\vn_{0}}}^{2}\Gamma$ is the magnitude of the coupling between mode $i$ and the reservoir, and the dynamical matrix $A_{i,j}$ is given by
\begin{equation}
	A_{i,j} = \gd_{i,j}\gve_{i} - i e^{i\p{\varphi_{j}-\varphi_{i}}}\half \sqrt{\bar \Gamma_{i}\bar \Gamma_{j}}.
\end{equation}

%%%%%%%%%%%%%%%%%%%%%%%%%%%%%%%%%%%%%%%%%%%%%%%%
\pardash{Ensuring a unique steady state}
The dynamics described by \cref{eq:evolbe} are controlled by the complex  eigenvalues $\gl$ of the matrix $A$.
Note first that energy eigenmodes having a node at the drain site are completely unaffected by the dissipation, and thus yield $\gl = \gve_{i}$. 
Such dark modes are a generic feature of Hamiltonians possessing degenerate spectra, as one can construct a basis for each $M$-fold degenerate subspace consisting of a single ``bright" mode which couples to the bath, and $M-1$ uncoupled ``dark" modes.  

The coupling to the reservoir will both mix and cause decay of the bright energy eigenmodes.  The resulting dynamical matrix eigenvalues can be written as ${\gl = \nu -i\half[\gamma]}$, where $\nu, \gamma$ are real solutions of the equation (see \cref{app:eigenvalues})
\begin{equation}\begin{gathered}
\sum\nolimits_{j}\frac{\half[\bar \Gamma_{j}]}{\half[\gamma] + i\p{\nu - \gve_{j}}} = 1. 
\label{eq:consist}
\end{gathered}\end{equation}
Examining the real part of this condition,  we immediately find that all bright eigenmodes have relaxation rates $\gamma > 0$.  Thus the ``bright" portion of the Hilbert space will relax to a unique steady state. In combination with the dark modes retaining their initial configuration this determines the system's final state.

A unique steady state is desirable for most applications of reservoir engineering.  There are two common sources of dark modes. Real space symmetries may result in a large number of modes having a node at high-symmetry sites  (e.g. anti-symmetric modes have nodes at the central site of a square lattice). These can be avoided by not placing the drain on these sites. A second source of dark modes is degeneracies in the spectrum, as discussed above. These can be removed by adding terms like an on-site potential which respects the system's chirality. We discuss this further below \cref{eq:Hsym}. 
Outside of these sources, a quadratic Hamiltonian will generically have few ($\ll \LL$) or no dark modes at any drain site, and the final state's correlation structure will be dominated by the engineered reservoir as long as their initial population is not too high.
In what follows, we assume that these conditions are met and there are no dark modes.  We will provide several concrete examples showing that these conditions are indeed achievable in realistic models.

%%%%%%%%%%%%%%%%%%%%%%%%%%%%%%%%%%%%%
\pardash{Chiral symmetry and the steady state}While the absence of dark modes ensures a unique steady state, it does not ensure that this state will be pure 
(see \cref{app:pure}). We find, however, that we can guarantee a pure steady state by imposing a simple symmetry requirement on our system: the existence of a generalized chiral symmetry which leaves the drain site invariant.  More explicitly, we require the eigenmodes of $\hat{\mathcal{H}}$ to come in pairs of opposite energy and equal wavefunction amplitude at the drain,
\begin{equation}
	\gve_{-i} = - \gve_{i}, \qquad \abs{\psi_{-i}\br{\vn_{0}}} = \abs{\psi_{i}\br{\vn_{0}}}.
\label{eq:symcond}
\end{equation}
Here we have indexed the modes ${i \in \{-\half[N],...,\half[N] \}}$. 

The above spectral structure arises in a large variety of tight-binding models, including disordered systems, and is often associated with a sublattice symmetry. For example, in the absence of any on-site potential, it is present in a 1D lattice with (arbitrary, possibly random) nearest neighbor hopping, and more generally in any system with a bipartite hopping structure \cite{Chalker2003}.  Other examples include the SSH model in 1D \cite{Heeger1988}, and in two dimensions graphene band structure \cite{Aoki2014b} and the Hofstadter model \cite{Hofstadter1976}. 

To understand how the chiral structure in \cref{eq:symcond} constrains the steady state, we first note that this structure ensures that $\hat{\mathcal{H}}$ is invariant under any two-mode squeezing (or Bogoliubov) transformation that mixes eigenmode operators $\hat{b}_{i}$ and $\hat{b}_{-i}\dg$.  
We thus define a new set of canonical annihilation operators ${\hat \gb_{i} = \cosh r \, \hat b_{i} - e^{i\p{\phi - \varphi_{i} - \varphi_{-i}}}\sinh r \, \hat b_{-i}\dg}$.  Note that the definition of these modes depends both on the properties of the squeezed reservoir (through $r$ and $\phi$), and on the position of the drain (through $\varphi_i$ and $\varphi_{-i}$).

Using \cref{eq:evolbe}, we find these new quasiparticle operators obey the equations of motion
\begin{equation}\begin{split}
	\dot{\hat \gb}_{i} & = -i \sum\nolimits_{j} A_{i,j}\hat\gb_{j} + e^{-i\varphi_{i}}\sqrt{\bar \Gamma_{i}}\hat \xi
	\\ & + \p{\abs{\psi_{i}\br{\vn_{0}}} - \abs{\psi_{-i}\br{\vn_{0}}}}\br{\p{\dotsi}\hat \gb_{j}\dg + \p{\dotsi}\hat \xi\dg + \dotsi}.
\label{eq:evolsym}
\end{split}\end{equation}
Here, ${\hat\xi = \cosh r \hat \zeta - e^{i\phi}\sinh r \hat\zeta\dg}$ is a noise operator with correlation functions corresponding to simple (unsqueezed)  vacuum noise:  ${\avg{\hat\xi\p{t}\hat\xi\dg\p{t\pr}} = \gd\p{t-t\pr}}$, ${\avg{\hat\xi\dg\p{t}\hat\xi\p{t\pr}} = \avg{\hat\xi\p{t}\hat\xi\p{t\pr}} = 0}$. 

The invariance of the drain site under the generalized chiral symmetry (c.f.~\cref{eq:symcond}) ensures that the second line of \cref{eq:evolsym} vanishes.  We thus find that the new $\hat{\beta}_{i}$ modes evolve with the same dynamical matrix as the original modes, but the noise term now corresponds to simple vacuum noise.  As the dynamical matrix is unchanged, we again have no dark modes.  Further, as the dynamical matrix corresponds to simple hopping and local damping, the unique steady state is the joint vacuum of all $\hat\gb_{i}$ modes.  This steady state yields non-trivial correlations between the original energy eigenmodes:  
\begin{equation}\begin{gathered}
	\avg{\hat b_{i}\dg\hat b_{j}} = \gd_{i,j} \Ns, \qquad \avg{\hat b_{i}\hat b_{j}} = \gd_{i,-j}e^{-i\p{\varphi_{i}+\varphi_{-i}}} \Ms.
	\label{eq:ssEcorrs}
\end{gathered}\end{equation}
Thus, the single, locally-coupled squeezed reservoir leads to a pure steady state where each pair of $\varepsilon, -\varepsilon$ energy eigenmodes is in a pure two-mode squeezed state with squeezing parameter $r$.

In real space, the steady state has a uniform average photon number on each site, an absence of any beam-splitter correlations, and non-trivial pattern of anomalous correlators:
\begin{equation}\begin{gathered}
	\avg{\hat a_{\vm}\dg\hat a_{\vn}} =  \gd_{\vm,\vn} \Ns,
	\qquad \avg{\hat a_{\vm}\hat a_{\vn}} =  \gs_{\vm,\vn} \Ms,
	\\ \gs_{\vm,\vn} = \sum\nolimits_{j}e^{-i\p{\varphi_{j}+\varphi_{-j}}}\psi_{j}\br{\vn}\psi_{-j}\br{\vm}.
\label{eq:ssrealcorrs}
\end{gathered}\end{equation}
The pattern of correlations depends explicitly on the position of the drain site via the phases $\varphi_j$. One always finds that ${\gs_{\vm,\vn_{0}} = \gd_{\vm,\vn_{0}}}$, implying that the drain site is in a pure squeezed state and unentangled with the rest of the lattice.

%%%%%%%%%%%%%%%%%%%%%%%%%%%%%%
\pardash{Connection to symmetry operations}
We see that all the non-trivial correlation structure of the pure steady state is contained in the matrix $\sigma$. While its formal definition in terms of eigenmodes may seem opaque, it has a simple physical meaning:  it directly defines a symmetry operation on $\H$.  More explicitly, the real-space tight-binding Hamiltonian matrix $H$ (c.f.~Eq.~(\ref{eq:HS})) satisfies (see \cref{app:matsymmetry}):
\begin{equation}
	\gs\dg\cdot H \cdot \gs = -H^{*}, \qquad \gs_{\vm,\vn_{0}} = \gd_{\vm,\vn_{0}}.
\label{eq:Hsym}
\end{equation}

The unitary (and symmetric) matrix $\sigma$ thus maps $H$ to $-H^*$.  At the level of operators, 
this equation can be expressed as
\begin{equation}\begin{gathered}
	\hat{\mathcal U}\hat \H + \hat \H \hat{\mathcal U} = 0.
	\label{eq:OpSymm}
\end{gathered}\end{equation}
The operator $\hat{\mathcal U}$ is most easily understood in the case where $\H$ has time reversal symmetry, such that we can work in a gauge where $H = H^{*}$, and where the eigenmode wavefunctions are real.  In this case, $\hat{\mathcal U}$ is a unitary symmetry operator associated with the chiral symmetry of the Hamiltonian,
\begin{equation}
\hat{\mathcal U} \to \hat{\mathcal S}: \qquad \hat{\mathcal S} \hat a_{\vm}\hat{\mathcal S}^{-1} = \sum\nolimits_{\vn}\gs_{\vm,\vn}\hat a_{\vn}.
	\label{eq:chiSymm}
\end{equation}
and the anomalous correlators in real space {\it are simply the matrix elements of the chiral symmetry operator}.

In the more general case where time-reversal symmetry is broken, the relevant symmetry is a particle-hole transformation:  
\begin{equation}
	\hat{\mathcal U} \to \hat{\mathcal C}: \quad \hat{\mathcal C} \hat a_{\vm}\hat{\mathcal C}^{-1} 
		= \sum\nolimits_{\vn}\gs_{\vm,\vn}\hat a_{\vn}\dg.
		\label{eq:ehSymm}
\end{equation}
We confirm that this definition satisfies \cref{eq:OpSymm} in \cref{app:opsymmetry}.
We again have that the pattern of anomalous correlations in the steady state is \emph{directly set by the real-space matrix elements of the particle-hole symmetry operator}.  

The upshot of our analysis is that the generalized chiral structure of $\H$ does more than ensure a pure steady state:  the corresponding symmetry operation directly determines its pattern of correlations.  Thus, one does not need knowledge of all the eigenmodes to understand the steady state, and simply identifying the relevant symmetry operation is enough.  This provides a powerful and very general principle for engineering steady states with the desired correlation patterns. Because it is linear, \cref{eq:Hsym} also allows us to modify the system while retaining chiral symmetry. For instance, any diagonal matrix which obeys \cref{eq:Hsym} represents a set of on-site potentials that can be added to the system without changing its steady state. Such terms can be useful in removing unwanted dark modes.

We now go on to provide several examples of systems with a generalized chiral symmetry.

%%%%%%%%%%%%%%%%%%%%%%%%%%%%%%%%%%%%%%%%%%%%%%%%%
\pardash{Chiral symmetry from bipartite hopping} Consider first a system with time-reversal symmetry, vanishing on-site energies and hopping terms that connect two distinct sub-lattices, with Hamiltonian
\begin{equation}
	\hat \H = \sum_{\mathclap{\vec a\in A,\vec b\in B}}J_{\vec a,\vec b}\p{\hat a_{\vec a}\dg\hat a_{\vec b} + \hat a_{\vec b}\dg\hat a_{\vec a}},
\end{equation}
for a real $J_{\vec a,\vec b}$ and some partition of the sites, $A\cap B = \varnothing$. The system has a unitary chiral symmetry defined by
\begin{equation}
\gs_{\vm,\vn} = \p{-1}^{s_{\vn}}\gd_{\vm,\vn},
\end{equation}
where $s_{\vn} = 0,1$ for $\vn\in A, B$ respectively.  A variety of tight-binding models have this form, including all bipartite lattices with nearest neighbor hopping (e.g.~a 1D chain or 2D square lattice); the symmetry holds with arbitrary (possibly random) matrix elements.  

It follows from \cref{eq:ssrealcorrs} that if we now locally couple this system to squeezed dissipation (with arbitrary choice of drain site), it will relax into a product state where each site is in a pure squeezed state with parameters $r,\phi$.  We thus have a robust method for preparing a lattice of squeezed states, using a single squeezing source.  Moreover, the steady state is robust against any amount of disorder in the hopping parameters.

\pardash{Generalized chiral symmetry from spatial inversion} 
Consider next a system where spatial inversion about the origin takes $H\to -H^{*}$.  With an appropriate labelling of lattice sites, this symmetry corresponds to
\begin{equation}
	J_{-\vn,-\vm} = -J_{\vm,\vn} \qquad V_{-\vn} = -V_{\vn}.
\end{equation}
Such systems formally have a particle-hole symmetry with a symmetry matrix
\begin{equation}\begin{gathered}
	\gs_{\vm,\vn} =  \gd_{\vm,-\vn}.
\end{gathered}\end{equation}
Note that inversion necessarily leaves the origin $\vn = \vec 0$ invariant.  Thus, if we couple the origin to our squeezed reservoir, 
$\vn_{0} = \vec 0$, the steady state is described by \cref{eq:ssrealcorrs} with $\gs_{\vm,\vn}$ given above.  The state thus factorizes into a product of two-mode squeezed states, with each site $\vn$ entangled with the site $-\vn$.  

A similar result is seen in a bipartite lattice with nearest neighbor hopping if the onsite energies are odd under inversion while the hoppings satisfy ${J_{-\vn,-\vm} = J_{\vm,\vn}}$. Coupling the drain at the origin again leads to pure two-mode squeezing described by ${\gs_{\vm,\vn} =  \p{-1}^{s_{\vn}} \gd_{\vm,-\vn}}$.  In the case of a 1D chain with uniform hopping, this corresponds to the model of Ref.~\onlinecite{Zippilli2015}. 

%%%%%%%%%%%%%%
\begin{figure}[t] %  figure placement: here, top, bottom, or page
   \centering
   \hspace{0.05\columnwidth} \scalebox{1.3}{$\abs{\avg{\hat a_{\vm}\hat a_{\vn}}}/\cosh r \sinh r$}   \hfill \includegraphics[height=\baselineskip]{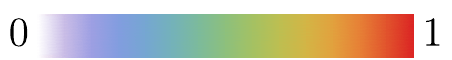} \hfill

   \subfloat[$\vn_{0} = \p{0,2}$]{\includegraphics[width=0.23\columnwidth]{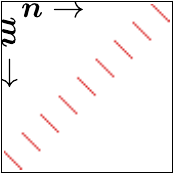}\label{fig:2D0yAll}}\hfill
   \subfloat[$\p{2,0}$]{\includegraphics[width=0.23\columnwidth]{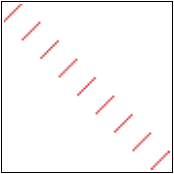}\label{fig:2Dx0All}}\hfill
   \subfloat[$\p{2,2}$]{\includegraphics[width=0.23\columnwidth]{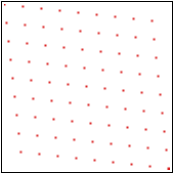}\label{fig:2DzzAll}}\hfill
   \subfloat[$\p{2,4}$]{\includegraphics[width=0.23\columnwidth]{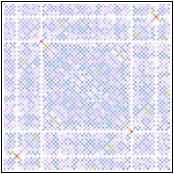}\label{fig:2DxyAll}}
   
   \vspace{\baselineskip}
   \hspace{0.05\columnwidth} \scalebox{1.1}{$\abs{\avg{\hat a_{\p{4,1}}\hat a_{\p{x,y}}}}/\cosh r \sinh r$}  \hfill\hfill
   \vspace{-0.5\baselineskip}
   
   %wLeg should be *203/188
   \subfloat[$\p{0,2}$]{\includegraphics[width=0.248\columnwidth]{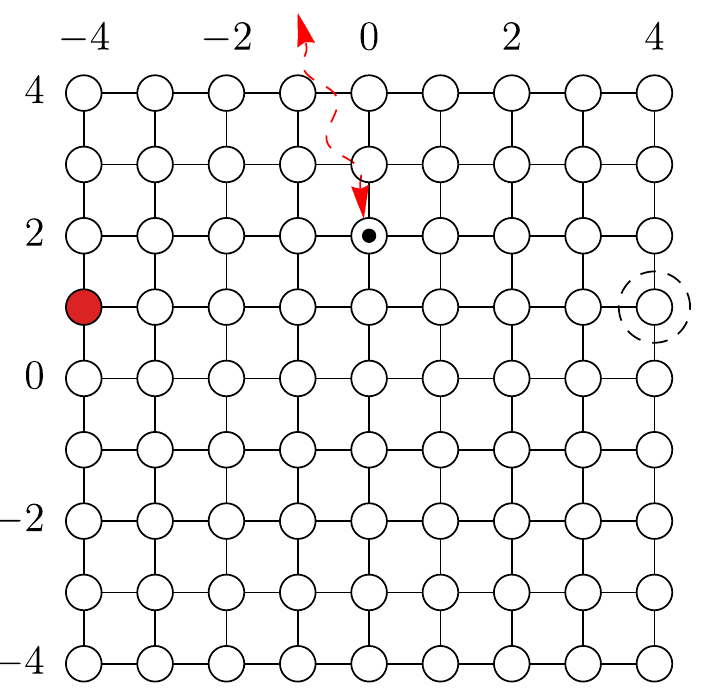}\label{fig:2D0y41}}\hfill
   \subfloat[$\p{2,0}$]{\includegraphics[width=0.23\columnwidth]{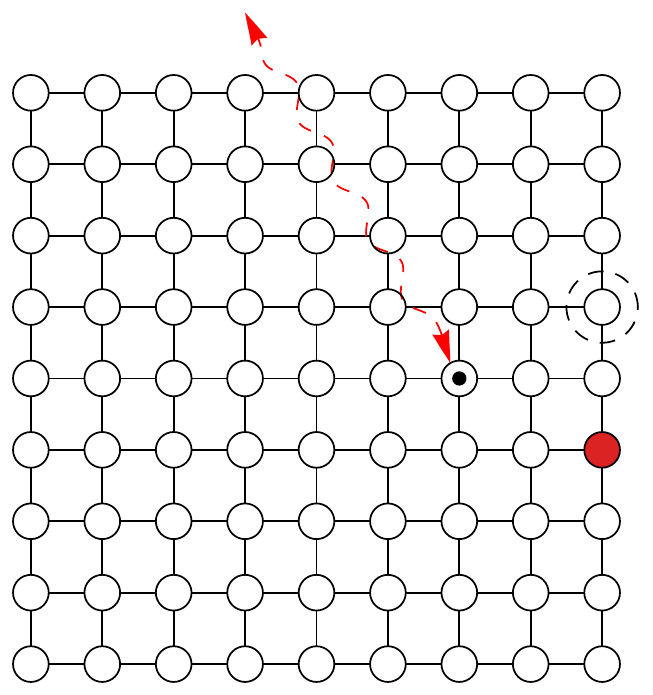}\label{fig:2Dx041}}\hfill
   \subfloat[$\p{2,2}$]{\includegraphics[width=0.23\columnwidth]{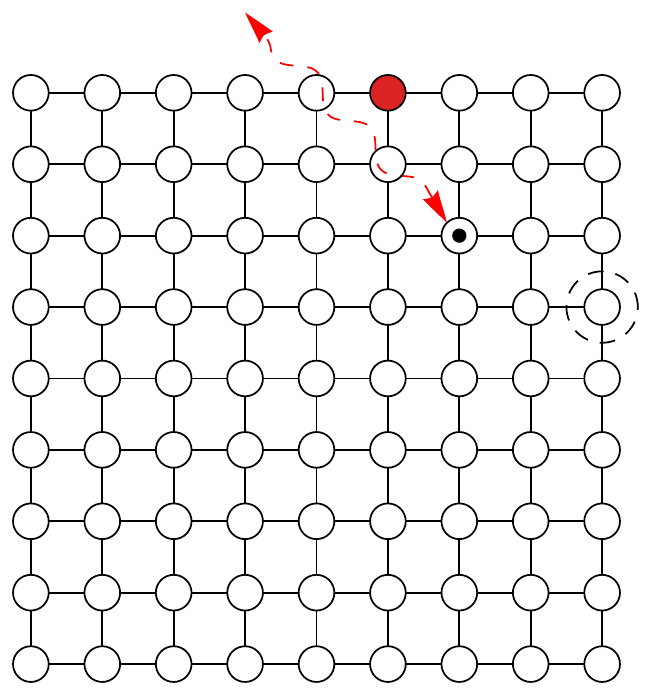}\label{fig:2Dzz41}}\hfill
   \subfloat[$\p{2,4}$]{\includegraphics[width=0.23\columnwidth]{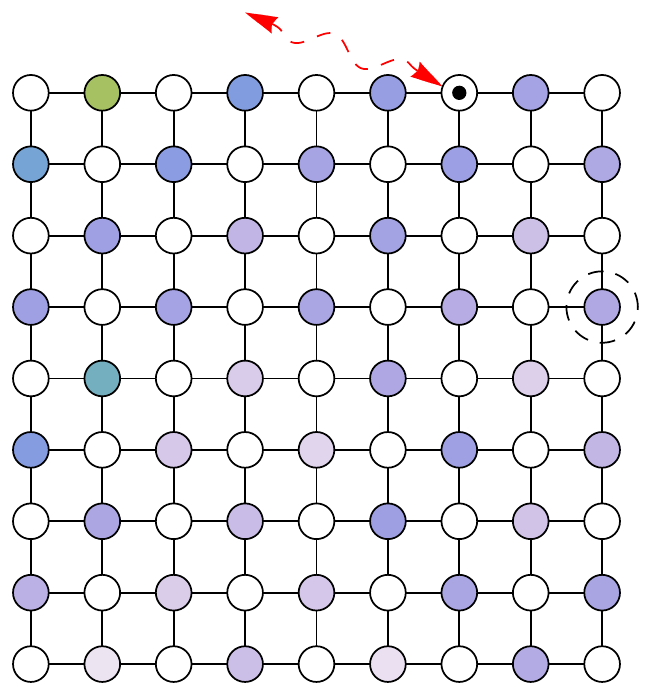}\label{fig:2Dxy41}}

   \caption{Steady state correlation patterns in a $9\times9$ Hofstadter lattice with plaquette flux $\Phi = \half[\pi]$, for various choices of the drain site  $\vec{n}_0$.  
    In all cases there are no dark modes, and hence a unique pure steady state.  
 \protect\subref{fig:2D0yAll}-\protect\subref{fig:2DxyAll}: The full $81\times 81$ matrix of anomalous correlations $\langle \hat{a}_{\vec m} \hat{a}_{\vec n} \rangle$, with the axes corresponding to an (arbitrary) labelling of the 81 lattice sites. 
 \protect\subref{fig:2D0y41}-\protect\subref{fig:2Dxy41}: Spatial pattern of correlations between site ${\vec m = \p{4,1}}$ (marked by a dashed circle) and every site on the lattice.
  Each panel corresponds to a different choice of drain site.  
As discussed in the main text, when $\vec{n}_0$ corresponds to a position of spatial symmetry, each site is correlated with just one other site. For other choices a complex pattern correlations emerges, as shown in \protect\subref{fig:2DxyAll},\protect\subref{fig:2Dxy41}.
 }
   \label{fig:2Dlat}
\end{figure}

%%%%%%%%%%%%%%%%%%%%%%

\pardash{Particle-hole symmetries in the Hofstadter model} Consider a two-dimensional square lattice with a (synthetic) flux $\Phi$ through each plaquette.  Labelling the sites via $\vn = (x,y)$ with $x,y \in (-M,M)$, we have: 
\begin{equation}
\hat \H = -J\sum_{\mathclap{x,y}}\p{\hat a_{\p{x+1,y}}\dg\hat a_{\p{x,y}} + e^{i\Phi x}\hat a_{\p{x,y+1}}\dg\hat a_{\p{x,y}} + \hc}.
\label{eq:H2Dlat}
\end{equation}
Such a system has recently been realized both with coupled optical cavities \cite{Hafezi2013}, and with coupled superconducting microwave cavities \cite{Anderson2016,Owens2017}.  One can in general choose $\Phi$ and $\vec n_{0}$ to ensure the absence of dark modes.  

For any value of the flux, this system has a  chiral symmetry $\hat{\mathcal S}$ described by 
${\gs_{\p{x,y},\p{x\pr,y\pr}} = \p{-1}^{x+y}} \delta_{x,x'}\delta_{y,y'}$.  The particle-hole symmetry $\C$ that is relevant to the steady-state entanglement 
depends on the choice of drain site and is generically nontritival.
If one places the drain site at certain positions with high spatial symmetry, it takes a simple form,
\begin{equation}\begin{split}
& \gs_{\p{x,y},\p{x\pr,y\pr}} = 
 \\& \quad \left\{ \begin{array}{lcl}\p{-1}^{x+y}\gd_{x,x\pr}\gd_{y,-y\pr} & & \vec n_{0} = \p{z,0}
 	\\ \p{-1}^{x+y}\gd_{x,-x\pr}\gd_{y,y\pr} & & \vec n_{0} = \p{0,z}
	\\ \p{-1}^{x+y}\gd_{x,y\pr}\gd_{y,x\pr}e^{i\Phi xy} & & \vec n_{0} = \p{z,z},\end{array}\right.
\label{eq:gsHof} 
\end{split}\end{equation}
where $z\ne 0$ is arbitrary
\footnote{Note that putting the drain site in the middle of the lattice, i.e.~ at $\p{0,0}$, always results in a large number of dark modes and is thus not an interesting case}.
For these positions, the steady state is a product of two-mode squeezed states in real space; the nature of the pairing depends on the choice of drain site.  These relatively simple steady states are shown in \cref{fig:2Dlat}.  

We stress that all of the transformations $\gs$ in \cref{eq:gsHof}  are symmetries of the system, and they are closely related: in terms of energy eigenstates, they all have the form ${\C\hat b_{i}\C^{-1} = e^{-i\p{\varphi_{i}+\varphi_{-i}} }\hat b_{-i}\dg}$. The choice of $\vn_{0}$ determines which is relevant for calculating steady state correlations. We discuss this further in \cref{app:2Dsym}. 

If we couple the reservoir at a different site, we still obtain a pure steady state, but one that does not factor simply in real space.
These states are reminiscent of complex multi-mode Gaussian entangled states known as cluster states (see \cref{app:cluster}). One example is shown in \cref{fig:2DxyAll,fig:2Dxy41}.

\begin{figure}[tbh] %  figure placement: here, top, bottom, or page
   \centering

   \includegraphics[width=0.8\columnwidth]{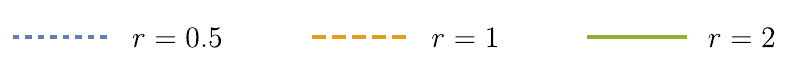}\vspace{-\baselineskip}

   %width = 1.185:1:1
   \subfloat[\scriptsize{With local disorder}]{\includegraphics[width=0.52\columnwidth]{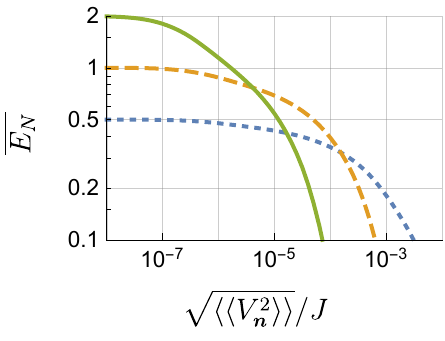} \label{fig:2Ddisorder}}
   \subfloat[\scriptsize{With internal loss}]{\includegraphics[width=0.44\columnwidth]{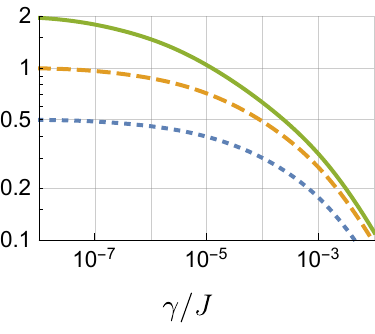} \label{fig:2Ddissip}}
%   \subfloat[\scriptsize{Time evolution}]{\includegraphics[width=0.30\columnwidth]{2DTimeEnt} \label{fig:2Dtime}}

   \caption{Robustness of the correlation structure depicted in \cref{fig:2Dlat}. 
   We  plot the entanglement per mirrored pair, ${\overline{E_{N}} = \tfrac{\ln \sqrt{2}}{\LL-\sqrt{\LL}}\sum_{x\ne y}E_{N}\br{\p{x,y},\p{y,x}}}$, 
   where $E_{N}[\vec{m},\vec{n}]$ is the log negativity of the two sites \cite{Vidal2001}. 
   All results are for a $9\times 9$ lattice with $\Phi = \half[\pi]$, $\vn_{0} = \p{2,2}$, and $\Gamma = 3J$. 
   	\protect\subref{fig:2Ddisorder} Decay of  ${\overline{E_{N}}} $ due to disorder:  we add a uniformly distributed random potential at each site, with $\avg{\avg{V_{\vn}}} = V_{\vn_{0}} = 0$ and variance as shown. The entanglement is averaged over 20 realizations of $V_{\vn}$. 
	 \protect\subref{fig:2Ddissip} Decay of ${\overline{E_{N}}}$ due to internal loss: we add loss at rate $\gamma$ to each site.  
%	 {\color{red}\protect\subref{fig:2Dtime} Evolution of the system towards its steady state, plotted in the presence of an internal loss $\gamma \sim 10^{-3}J$. }
	 }
   \label{fig:2Ddis}
\end{figure}

\pardash{Robustness of the steady state} Our discussion so far has assumed perfect chirality in the Hamiltonian. In \cref{fig:2Ddis} we assess the robustness of the protocol against experimental limitations which break this chirality. We calculate how local disorder and internal loss reduce the amount of entanglement in the steady state.

Our protocol's advantage compared with more elaborate setups (e.g.~\cite{Diehl2008}) is in its experimental simplicity. The use of a single dissipator, however, increases the relaxation time of the system.
%, which will scale as $t_{\rm relax} > N/\Gamma$. 
This leaves the steady state vulnerable to any processes that occur on a shorter time scale. \Cref{fig:2Ddissip} shows the result of this interplay. We find that a large amount of entanglement survives even when these processes occur with rates of 0.1\%-1\% of the hopping rate, an experimentally realistic level in superconducting cavities \footnote{D. Schuster, private communication} and optomechanics \cite{Reinhardt2015}.

%%%%%%%%%%%%%
%%%%%%%%%%%%%%

%%%%%%%%%%%
\pardash{Conclusion}
We have developed a simple yet potentially powerful symmetry-based approach for reservoir engineering of entangled steady states of a bosonic lattice.  Our approach only employs a single, locally coupled squeezed reservoir, and relies on the existence of chiral symmetry, something that is present in a variety of different tight-binding models.  The approach allows the preparation and stabilization of a variety of different kinds of Gaussian entangled states, and is applicable to current state-of-the-art quantum photonic systems.  In particular, experiments in superconducting quantum circuits have demonstrated all the required ingredients, including the realization of a Hofstadter model \cite{Owens2017}, and the ability to strongly couple to a squeezed vacuum reservoir  \cite{Flurin2012,Murch2013,Toyli2016,Clark2016}.  Our work also suggests that symmetry-based approaches could be a powerful way to design lattice reservoir engineering protocols in more complex systems (e.g.~where interactions and nonlinearity are also important).

\begin{acknowledgments}
This work was supported by NSERC.
\end{acknowledgments}

\newpage
\appendix
\begin{widetext}

\section{Derivation of the dissipation spectrum\label{app:eigenvalues}}

The evolution equations for the eigenmodes of the system are
\begin{equation}\begin{gathered}
\dot{\hat b}_{i} = -i\sum\nolimits_{j} A_{i,j}\hat b_{j} + e^{-i\varphi_{i}}\sqrt{\bar \Gamma_{i}}\hat\zeta,
\\  A_{i,j} = \gd_{i,j}\gve_{i} - i e^{i\p{\varphi_{j}-\varphi_{i}}}\half \sqrt{\bar \Gamma_{i}\bar \Gamma_{j}}, 
\\ \bar \Gamma_{i} = \abs{\psi_{i}\br{\vn_{0}}}^{2}\Gamma, \qquad \varphi_{i} = \arg\p{\psi_{i}\br{\vn_{0}}}.
\end{gathered}\end{equation}

We seek the eigenmodes of $A$ and their eigenvalues. These have
\begin{equation}
\tilde b_{\nu} = \sum_{j}u_{\nu,j}\hat b_{j} \qquad \dot{\tilde b}_{\nu} = -\p{\half[\gamma] + i\nu}\tilde b_{i} + g_\nu\hat\zeta.
\end{equation}
Calculating,
\begin{equation*}\begin{gathered}
\begin{split}\dot{\tilde b}_{\nu} & = \sum_{j}u_{\nu,j}\br{-i\gve_{j}\hat b_{j} + e^{-i\varphi_{j}}\sqrt{\bar\Gamma_{j}}\p{-\half\sum_{l}e^{i\varphi_{l}}\sqrt{\bar \Gamma_{l}}\hat b_{l} + \hat\zeta}}
	\\ & = -\sum_{j}\p{\half g_\nu e^{i\varphi_{j}}\sqrt{\bar\Gamma_{j}}+i\gve_{j}u_{\nu,j}}\hat b_{j} + g_{\nu}\hat\zeta \qquad g_{\nu} = \sum_{j}u_{\nu,j}e^{-i\varphi_{j}}\sqrt{\bar \Gamma_{j}}
\\ \dot{\tilde b}_{\nu} & = -\p{\half[\gamma] + i\nu}\tilde b_{\nu} + g_\nu\hat\zeta = -\sum_{j}\p{\half[\gamma] + i\nu}u_{\nu,j}\tilde b_{j} + g_\nu\hat\zeta\end{split}
\\ \Rightarrow  \br{\half[\gamma] + i\p{\nu - \gve_{j}}}u_{\nu,j} = \half g_{\nu} e^{i\varphi_{j}}\sqrt{\bar\Gamma_{j}}
\end{gathered}\end{equation*}

\begin{itemize}
\item A solution with $\gamma = 0$, $\nu = \gve_{i}$ for some $i$ is consistent only if $\bar \Gamma_{i} = 0$, with $u_{\nu,j} = \gd_{i,j}$ and $g_{\nu} = 0$. These are the dark modes, which are unaffected by the dissipation.
\item Otherwise, we solve for $u_{\nu,j}$ and find the self-consistency equation:
\begin{equation}
g_{\nu} = \sum_{j}u_{\nu,j}e^{-i\varphi_{j}}\sqrt{\bar \Gamma_{j}} 
	= \sum_{j}\tfrac{g_{\nu}\half[\sqrt{\bar \Gamma_{j}}]}{\half[\gamma] + i\p{\nu - \gve_{j}}}\sqrt{\bar \Gamma_{j}} 
	= g_{\nu}\sum_{j}\tfrac{\half[\bar \Gamma_{j}]}{\half[\gamma] + i\p{\nu - \gve_{j}}}.
\end{equation}
\end{itemize}

\section{Non-pure Steady States from Zero-Entropy Reservoir\label{app:pure}}

We note in the main text, below \cref{eq:consist}, the requirement for a unique steady state is that there are no dark states of the dissipator. We note that this guarantees a unique state but not a unique pure state, even for a zero-entropy reservoir such as the one we use. To demonstrate this, consider a two-mode syste,
\begin{equation}
\hat \H = \half[V]\p{\hat a_{1}\dg\hat a_{1} - \hat a_{2}\dg\hat a_{2}} - J\p{\hat a_{1}\dg\hat a_{2} + \hat a_{2}\dg\hat a_{1}}.
\end{equation}
Choosing site $n_{0} = 1$ as the drain, we have the steady state given by
\begin{equation}\begin{gathered}
\nonumber
\avg{\hat a_{1}\dg\hat a_{1}} = \avg{\hat a_{2}\dg\hat a_{2}} = \sinh^{2}r \qquad \avg{\hat a_{1}\dg\hat a_{2}} = 0
\\ \avg{\hat a_{1}\hat a_{1}} = \tfrac{4J^{2} - \Gamma V}{4J^{2} + V^{2} - i \Gamma V} \qquad \avg{\hat a_{2}\hat a_{2}} = \tfrac{-4J^{2}}{4J^{2} + V^{2} - i \Gamma V}
	\qquad \avg{\hat a_{1}\hat a_{2}} =  \tfrac{2JV}{4J^{2} + V^{2} - i \Gamma V}
\end{gathered}\end{equation}
and we can calculate the purity of the state,
\begin{equation}
\mu = \sqrt{\tfrac{\p{4J^{2}+V^{2}}^{2} + \Gamma^{2}V^{2}}{\p{4J^{2}+V^{2}\cosh^{2}2r}^{2} + \Gamma^{2}V^{2}\cosh^{2}2r}}.
\end{equation}
As expected, it is a pure state when $V = 0$, and we have a chiral system with $\abs{\psi_{+}\br{1}} = \abs{\psi_{-}\br{1}}$, or when $r = 0$, and there is no squeezing. Otherwise, the steady state is a mixed state.

\section{Relation of the symmetry to the correlation matrix\label{app:matsymmetry}}

The relation between the eigenmodes and the original basis is
\begin{equation}
\hat b_{i} = \sum_{\vn}\p{\psi_{i}\br{\vn}}^{*}\hat a_{\vn} \qquad \hat a_{\vn} = \sum_{j}\psi_{j}\br{\vn}\hat b_{j}
\end{equation}
where the wavefunctions are orthonormal,
\begin{equation}
\sum_{\vn}\p{\psi_{i}\br{\vn}}^{*}\psi_{j}\br{\vn} = \gd_{i,j} \qquad \sum_{i}\p{\psi_{i}\br{\vn}}^{*}\psi_{i}\br{\vec m} = \gd_{\vn,\vec m}.
\end{equation}

The Hamiltonian can be written in the form
\begin{equation}
\hat \H = \sum_{\vec m,\vn}H_{\vec m,\vn}\hat a_{\vec m}\dg\hat a_{\vn} = \sum_{i}\gve_{i}\hat b_{i}\dg\hat b_{i}
\end{equation}
with the matrix relations
\begin{equation}
H_{\vec m,\vn} = \sum_{i}\psi_{i}\br{\vec m}\gve_{i}\p{\psi_{i}\br{\vn}}^{*}.
\end{equation}

In the eigenmode basis, we have at the steady state
\begin{equation}
\avg{\hat b_{i}\dg\hat b_{j}} \to \gd_{i,j}N, \qquad \avg{\hat b_{i}\hat b_{j}} \to \gd_{i,-j}e^{-i\p{\varphi_{i}+\varphi_{-i}}} M.
\end{equation}
and therefore, in real space
\begin{equation}\begin{gathered}
\avg{\hat a_{\vm}\dg\hat a_{\vn}} = \sum_{i,j}\p{\psi_{i}\br{\vm}}^{*}\psi_{j}\br{\vn}\avg{\hat b_{i}\dg\hat b_{j}} \to \sum_{i}\p{\psi_{i}\br{\vm}}^{*}\psi_{i}\br{\vn} N = \gd_{\vm,\vn}N
\end{gathered}\end{equation}
\begin{equation}\begin{gathered}
\avg{\hat a_{\vm}\hat a_{\vn}} =  \sum_{i,j}\psi_{i}\br{\vm}\psi_{j}\br{\vn}\avg{\hat b_{i}\hat b_{j}} \to  \sum_{i}\psi_{i}\br{\vm}\psi_{-i}\br{\vn}e^{-i\p{\varphi_{i}+\varphi_{-i}}}M \equiv \gs_{\vm,\vn}M
\end{gathered}\end{equation}

We then observe
\begin{equation}\begin{split}
\nonumber
& \p{\gs\dg\cdot H \cdot \gs}_{\vm,\vn}  = \sum_{\vec a,\vec b}\gs_{\vec a,\vm}^{*}H_{\vec a,\vec b}\gs_{\vec b,\vn}
	\\ & = \sum_{\vec a,\vec b}\mat{\sum_{i}e^{-i\p{\varphi_{i}+\varphi_{-i}}}\psi_{i}\br{\vec a}\psi_{-i}\br{\vm}}^{*}\mat{\sum_{j}\psi_{j}\br{\vec a}\gve_{j}\p{\psi_{j}\br{\vec b}}^{*}}
		\mat{\sum_{l}e^{-i\p{\varphi_{l}+\varphi_{-l}}}\psi_{l}\br{\vec b}\psi_{-l}\br{\vn}}
	\\ & = \sum_{i,j,l}e^{i\p{\varphi_{i} + \varphi_{-i} -\varphi_{l} - \varphi_{-l}}}\p{\psi_{-i}\br{\vm}}^{*}\mat{\sum_{\vec a}\p{\psi_{i}\br{\vec a}}^{*}\psi_{j}\br{\vec a}}
		\gve_{j}\mat{\sum_{\vec b}\p{\psi_{j}\br{\vec b}}^{*}\psi_{l}\br{\vec b}}\psi_{-l}\br{\vn}
	\\ & = \sum_{i,j,l}e^{i\p{\varphi_{i} + \varphi_{-i} -\varphi_{l} - \varphi_{-l}}}\p{\psi_{-i}\br{\vm}}^{*}\gd_{i,j}\gve_{j}\gd_{j,l}\psi_{-l}\br{\vn}
	\\ & = \sum_{i}\p{\psi_{-i}\br{\vm}}^{*}\gve_{i}\psi_{-i}\br{\vn} = \sum_{i}\p{\psi_{-i}\br{\vm}}^{*}\p{-\gve_{-i}}\psi_{-i}\br{\vn} = -H_{\vm,\vn}^{*}.
\end{split}\end{equation}

\section{Operator Formulation of Symmetry\label{app:opsymmetry}}

We show that the transformations of \cref{eq:chiSymm,eq:ehSymm} satisfy \cref{eq:Hsym}. The Hamiltonian is, again,
\begin{equation}
\hat \H = \sum_{\vec m,\vn}H_{\vec m,\vn}\hat a_{\vec m}\dg\hat a_{\vn}.
\end{equation}

In the real case, we have
\begin{equation}
\hat{\mathcal U} \to \hat{\mathcal S}: \qquad \hat{\mathcal S} \hat a_{\vm}\hat{\mathcal S}^{-1} = \sum\nolimits_{\vn}\gs_{\vm,\vn}\hat a_{\vn},
\end{equation}
and
\begin{equation}\begin{split}
\nonumber
\hat{\mathcal U}\hat \H\hat{\mathcal U}^{-1} & = \sum_{\vm,\vn}H_{\vm,\vn}\hS\hat a_{\vm}\dg\hat a_{\vn}\hS^{-1}
	 = \sum_{\vm,\vn}H_{\vm,\vn}\hS\hat a_{\vm}\dg \hS^{-1}\hS\hat a_{\vn}\hS^{-1}
\\& = \sum_{\vm,\vn}H_{\vm,\vn}\p{\sum_{\vm\pr}\gs\dg_{\vm\pr,\vm}\hat a_{\vm\pr}\dg }\p{\sum_{\vn\pr}\gs_{\vn,\vn\pr}\hat a_{\vn\pr}}
	= \sum_{\vm\pr,\vn\pr}\p{\gs\dg\cdot H\cdot\gs}_{\vm\pr,\vn\pr}\hat a_{\vm\pr}\dg\hat a_{\vn\pr} 
\\& = \sum_{\vm\pr,\vn\pr}-H^{*}_{\vm\pr,\vn\pr}\hat a_{\vm\pr}\dg\hat a_{\vn\pr}
	= - \sum_{\vm\pr,\vn\pr}H_{\vm\pr,\vn\pr}\hat a_{\vm\pr}\dg\hat a_{\vn\pr} = -\hat\H.
\end{split}\end{equation}

More generally,
\begin{equation}
\hat{\mathcal U} \to \C: \qquad \C \hat a_{\vm}\C^{-1} = \sum\nolimits_{\vn}\gs_{\vm,\vn}\hat a_{\vn}\dg,
\end{equation}
and
\begin{equation}\begin{split}
\nonumber
\hat{\mathcal U}\hat \H\hat{\mathcal U}^{-1} & = \sum_{\vm,\vn}H_{\vm,\vn}\C\hat a_{\vm}\dg\hat a_{\vn}\C^{-1}
	 = \sum_{\vm,\vn}H_{\vm,\vn}\C\hat a_{\vm}\dg \C^{-1}\C\hat a_{\vn}\C^{-1}
\\& = \sum_{\vm,\vn}H_{\vm,\vn}\p{\sum_{\vm\pr}\gs\dg_{\vm\pr,\vm}\hat a_{\vm\pr} }\p{\sum_{\vn\pr}\gs_{\vn,\vn\pr}\hat a_{\vn\pr}\dg}
	= \sum_{\vm\pr,\vn\pr}\p{\gs\dg\cdot H\cdot\gs}_{\vm\pr,\vn\pr}\hat a_{\vm\pr}\hat a_{\vn\pr} \dg
\\& = \Tr H + \sum_{\vm\pr,\vn\pr}-H\dg_{\vn\pr,\vm\pr}\hat a_{\vn\pr}\dg\hat a_{\vm\pr}
	=  - \sum_{\vm\pr,\vn\pr}H_{\vm\pr,\vn\pr}\hat a_{\vm\pr}\dg\hat a_{\vn\pr} = -\hat\H.
\end{split}\end{equation}
where $\Tr H = \Tr \p{-H} = 0$.

\section{Symmetry transformations in a 2D Lattice\label{app:2Dsym}}
In the text, we discuss the the Hofstadter Hamiltonian,
\begin{equation}
\hat \H = -J\sum_{\mathclap{x,y}}\p{\hat a_{\p{x+1,y}}\dg\hat a_{\p{x,y}} + e^{i\Phi x}\hat a_{\p{x,y+1}}\dg\hat a_{\p{x,y}}} + \hc
\end{equation}
and the three real-space transformations that should be used depending on the position of the drain site,
\begin{equation*}\begin{gathered}
\C \hat a_{\p{x,y}}\C^{-1} = \sum_{x\pr,y\pr}\gs_{\p{x,y},\p{x\pr,y\pr}}\hat a_{\p{x\pr,y\pr}}\dg
\\ \begin{split} \C_{z,0} & \;\; \Rightarrow \;\; \gs_{\p{x,y},\p{x\pr,y\pr}} = \p{-1}^{x+y}\gd_{x,-x\pr}\gd_{y,y\pr}
\\ \C_{0,z} &\;\; \Rightarrow \;\; \gs_{\p{x,y},\p{x\pr,y\pr}} = \p{-1}^{x+y}\gd_{x,x\pr}\gd_{y,-y\pr}
\\ \C_{z,z} & \;\; \Rightarrow \;\; \gs_{\p{x,y},\p{x\pr,y\pr}} = \p{-1}^{x+y}\gd_{x,y\pr}\gd_{y,x\pr}e^{i\Phi xy}.
\end{split}\end{gathered}\end{equation*}

It's important to stress that these transformations are different in real space, but very similar in their effects on the eigenmodes. To stress this, we show the effect of these transformations at $\Phi = 0$, where the model is exactly solvable. We note that in the absence of a flux, $\Phi$, the bright modes behave in much the same way as in its presence. However, the $\Phi = 0$ model is highly-degenerate and has many dark modes.

We take a 2D lattice of size $N = 2M+1\times 2M+1$. Its eigenmodes are given by
\begin{equation}
\hat a_{\p{k,q}} = \tfrac{1}{M+1}\sum_{x,y}\sin\br{k \p{x+M+1}}\sin\br{q\p{y+M+1}}\hat a_{\p{x,y}}
\end{equation}
for $k,q\in\tfrac{\pi}{2\p{M+1}}\times \acom{1,\dotsc, 2M+1}$, with energies
\begin{equation}
\br{\hat a_{\p{k,q}},\hat H} = -2J\p{\cos k + \cos q}\hat a_{\p{k,q}}.
\end{equation}
Note that $\gve_{\p{\pi-k,\pi-q}} = -\gve_{\p{k,q}}$ and that $\gve_{\p{q,k}} = \gve_{\p{k,q}}$.

We note first the chiral symmetry,
\begin{equation}
\hat{\mathcal S}\hat a_{\p{x,y}}\hat{\mathcal S}^{-1} = \sum_{x\pr,y\pr} \p{-1}^{x+y}\gd_{x,x\pr}\gd_{y,y\pr}\hat a_{\p{x\pr,y\pr}}
\end{equation}
has
\begin{equation*}\begin{split}
\hat{\mathcal S}\hat a_{\p{k,q}}\hat{\mathcal S}^{-1} & = \tfrac{1}{M+1}\sum_{x,y}\sin\br{k \p{x+M+1}}\sin\br{q\p{y+M+1}}\p{-1}^{x+y}\hat a_{\p{x,y}}
\\ & =  \tfrac{1}{M+1}\sum_{x,y}\sin\br{\p{\pi - k} \p{x+M+1}}\sin\br{\p{\pi -q}\p{y+M+1}}\hat a_{\p{x,y}}  = \hat a_{\p{\pi-k,\pi-q}}.
\end{split}\end{equation*}
This guarantees the existence of a chiral structure and therefore particle-hole symmetries.

Next, we examine the first two particle-hole symmetries,
\begin{equation*}\begin{split}
\C_{z,0}\hat a_{\p{k,q}}\C_{z,0}^{-1} & = \tfrac{1}{M+1}\sum_{x,y}\sin\br{k \p{x+M+1}}\sin\br{q\p{y+M+1}}\p{-1}^{x+y}\hat a_{\p{-x,y}}\dg
	\\ & = \tfrac{1}{M+1}\sum_{x,y}\sin\br{k \p{x-M-1}}\sin\br{\p{\pi-q}\p{y+M+1}}\p{-1}^{x+M+1}\hat a_{\p{x,y}}\dg
	\\ & = -\tfrac{\p{-1}^{k/\tfrac{\pi}{2\p{M+1}}}}{M+1}\sum_{x,y}\sin\br{\p{\pi - k} \p{x+M+1}}\sin\br{\p{\pi -q}\p{y+M+1}}\hat a_{\p{x,y}}\dg
	\\ & = -\p{-1}^{k/\tfrac{\pi}{2\p{M+1}}}\hat a_{\p{\pi-k,\pi-q}}\dg
\\ \C_{0,z}\hat a_{\p{k,q}}\C_{0,z}^{-1} & = -\p{-1}^{q/\tfrac{\pi}{2\p{M+1}}}\hat a_{\p{\pi-k,\pi-q}}\dg.
\end{split}\end{equation*}
They both take each mode, up to a phase, to the antiparticle of its negative energy mode.

Finally, the last symmetry operator has
\begin{equation}
\nonumber
\C_{z,z}\hat a_{\p{k,q}}\C_{z,z}^{-1} = \tfrac{1}{M+1}\sum_{x,y}\sin\br{k \p{x+M+1}}\sin\br{q\p{y+M+1}}\p{-1}^{x+y}\hat a_{\p{y,x}}\dg = \hat a_{\p{\pi-q,\pi-k}}\dg.
\end{equation}
It takes each mode to a \emph{different} mode anti-particle, still corresponding to the same negative energy. This is possible because the zero-flux mode has the degeneracies mentioned above. The addition of flux breaks this degeneracy.

\section{Connection to cluster states and $\mathcal H$-graph states\label{app:cluster}}

A cluster state is a highly-entangled state of a many-qubit \cite{Briegel2001} or many-oscillator \cite{Zhang2006} system. Their high degree of entanglement makes them useful resource in a variety of quantum computing and quantum communication applications.

A continuous-variable cluster state is given by the wavefunction \cite{Menicucci2011}
\begin{equation}
\ket{\psi_{A}} = \exp\br{\half[i] \sum_{m,n}A_{mn}\hat x_{m} \hat x_{n}}\vac
\end{equation}
where the real, symmetric matrix $A$ is the \emph{adjacency matrix} representing a graph of connections. It also has a set of nullifiers, given by $\p{\hat p_{m} - \sum_{n}A_{m,n}\hat x_{n}}\ket{\psi_{A}} = 0$.

A similarly powerful state, which can be generated by pumping an optical parametric oscillator \cite{Menicucci2007}, is known as an $\mathcal H$-graph state and takes the form
\begin{equation}\begin{split}
\ket{\psi_{Z}} & = \exp\br{-\tfrac{i}{2} \sum_{m,n} \ga G_{m,n}\p{\hat x_{m}\hat p_{n} + \hat p_{m}\hat x_{n}}}\vac
	\\ & = \exp\br{\tfrac{1}{2} \sum_{m,n} \ga G_{m,n}\p{\hat a_{m}\dg \hat a_{n}\dg   - \hat a_{m}\hat a_{n}}}\vac
\end{split}\end{equation}
where the adjacency matrix is now $Z = i\exp\br{-2\ga G}$ for some real $G$.

The steady state described by \cref{eq:ssEcorrs,eq:ssrealcorrs} in the main text, which is described by 
\begin{equation}\begin{split}
\ket{\psi_{\gs}} 
	= \exp\br{\half \sum_{\vm,\vn}r\p{ e^{i\phi}\gs_{\vm,\vn}\hat a_{\vm}\dg\hat a_{\vn}\dg - e^{-i\phi}\gs_{\vm,\vn}^{*}\hat a_{\vm}\hat a_{\vn}}}\vac,
\end{split}\end{equation}
can clearly be taken as an extension of the $\mathcal H$-graph state to a complex adjacency matrix.

We note also that it has a set of nullifier states, given by $\p{\hat p_{\vm} - \sum_{\vn}\tilde A_{\vm,\vn} \hat x_{\vn}}\ket{\psi_{\gs}}$
where 
\begin{equation}
\tilde A = \p{I + \tanh r \Re \br{e^{i\phi}\gs}}\cdot  \p{\tanh r \Im \br{e^{i\phi}\gs}}.
\end{equation}

\end{widetext}
\bibliographystyle{apsrev4-1}

%\bibliography{/Users/yarivyanay/Documents/Citations/library}

\end{document}